\documentstyle[twoside,fleqn,espcrc2,epsf]{article}

\newcommand{\AmS}{{\protect\the\textfont2 
  A\kern-.1667em\lower.5ex\hbox{M}\kern-.125emS}} 
 
\def\Preprint{\vspace*{-7.0cm}   
  \noindent FTUV/99-69 \\ 
  IFIC/99-72 \\  
  \vspace{4.4cm}}

\def\refjl#1#2#3#4#5#6{\bibitem{#1} #2, {\it #3} {\bf #4} (#5) #6.} 

\def\etal{{\it et al}} 
\def\NP{Nucl. Phys.} 
\def\NPPS{Nucl. Phys. B (Proc. Suppl.)} 
\def\PL{Phys. Lett.} 
\def\PRL{Phys. Rev. Lett.} 
\def\PR{Phys. Rev.} 
 
\def\ZP{Z. Phys.} 
\def\EJ{Eur. Phys. J.}

\def\RPP{Rep. Prog. Phys.}

\newcommand{\eqn}[1]{(\ref{#1})} 
\newcommand{\be}{\begin{equation}} 
\newcommand{\ee}{\end{equation}} 
\newcommand{\no}{\nonumber} 
\newcommand{\bel}[1]{\be\label{#1}} 
\newcommand{\ba}{\begin{array}{c}} 
\newcommand{\bat}{\begin{array}{cc}} 
\newcommand{\ea}{\end{array}} 
\newcommand{\beqn}{\begin{eqnarray}} 
\newcommand{\eeqn}{\end{eqnarray}} 
 
\newcommand{\bi}{\begin{itemize}} 
\newcommand{\ei}{\end{itemize}}

\newcommand{\rms}{\rm\scriptsize}

\newcommand{\cP}{{\cal P}} 
 
\newcommand{\cH}{{\cal H}} 
\newcommand{\cA}{{\cal A}}

\begin{document} 
 
\title{Electroweak Precision Tests 
 \thanks{Invited talk at the International Workshop 
 {\it Particles in Astrophysics and Cosmology: from Theory
 to Observation} (Val\`encia, 3--8 May 1999)} 
} 
 
\author{A. Pich\address{Departament de F\'{\i}sica Te\`orica,  
         IFIC, Universitat de Val\`encia -- CSIC, \\  
         Apt. Correus 2085, E--46071 Val\`encia, Spain}}
 
\begin{abstract} 
\noindent 
Precision measurements of electroweak observables provide 
stringent tests of the Standard Model structure and an 
accurate determination of its parameters. 
An overview of the present experimental status is presented. 
\end{abstract} 
 
\maketitle 
\Preprint 
 
\section{INTRODUCTION} 
\label{sec:introduction} 
 
The Standard Model (SM) constitutes one of the most successful achievements 
in modern physics. It provides a very elegant theoretical 
framework, which is able to describe all known experimental 
facts in particle physics. 
A detailed description of the SM 
and its impressive phenomenological success 
can be found in Refs.~\cite{jaca:94} 
and \cite{sorrento:94}, which discuss the electroweak and strong 
sectors, respectively. 
 
The high accuracy achieved by the most recent experiments 
allows to make stringent tests of the SM structure at the level 
of quantum corrections. 
The different measurements complement each other in their different 
sensitivity to the SM parameters.  
Confronting these measurements with the theoretical predictions, one 
can check the internal consistency of the SM framework and 
determine its parameters. 
 
The following sections provide an overview of our 
present experimental knowledge on the electroweak couplings. 
A brief description of some classical QED tests is presented 
in Section~\ref{sec:qed}. 
The leptonic couplings of the $W^\pm$ bosons are analyzed in 
Section~\ref{sec:cc}, where the tests on lepton universality 
and the Lorentz structure of the $l^-\to\nu_l l'^-\bar\nu_{l'}$ 
transition amplitudes are discussed. 
Section~\ref{sec:nc} describes the status of the 
neutral--current sector, using the latest experimental results reported by 
LEP and SLD. 
Some summarizing comments are finally given in Section~\ref{sec:summary}. 
%
 
\section{QED} 
\label{sec:qed}

The most stringent QED test 
\cite{KI:90,KI:96,CKM:96,PPR:95,KKSS:92,KR:97,LR96,dRA:94,HK:97,BPP:96,ADH:99}
comes from the high--precision 
measurements \cite{PDG:98} of the $e$ and $\mu$ anomalous magnetic moments  
$a_l^\gamma\equiv (g_l-2)/2$: 
\beqn\label{eq:a_e} 
a_e^\gamma & \!\!\! = &\!\!\!\left\{ \!\bat
(115 \, 965 \, 215.4\pm 2.4) \times 10^{-11} & (\mbox{\rm Theory}) 
\\  
(115 \, 965 \, 219.3\pm 1.0) \times 10^{-11} & (\mbox{\rm Exp.}) 
\ea   \right.\no\\ && \no\\ 
a_\mu^\gamma &\!\!\! = &\!\!\!\left\{ \!\bat 
(1 \, 165 \, 916.0\pm 0.7)    
\times 10^{-9} & (\mbox{\rm Theory}) 
\\ 
(1 \, 165 \, 923.0\pm 8.4) \times 10^{-9} & (\mbox{\rm Exp.}) 
\ea   \right. \no  
\eeqn 
The impressive agreement between theory and experiment  
(at the level of the ninth digit for $a_e^\gamma$) 
promotes QED 
to the level of the best theory ever build by the human mind 
to describe nature. 
Hypothetical {\it new--physics} effects are  
constrained to the ranges 
$|\delta a_e^\gamma | < 0.9 \times 10^{-10}$ and 
$|\delta a_\mu^\gamma | < 2.4 \times 10^{-8}$ (95\% CL). 
 
To a measurable level, $a_e^\gamma$ arises entirely from virtual electrons and 
photons; these contributions are known \cite{KI:96} to $O(\alpha^4)$.  
The sum of all other QED corrections, associated with higher--mass 
leptons or intermediate quarks, only amounts to  
$+(0.4366\pm 0.0042)\times 10^{-11}$, 
while the weak interaction effect is a tiny $+0.0030\times 10^{-11}$; 
these numbers \cite{KI:96} are well below the present  
experimental precision. 
The theoretical error is dominated by the uncertainty in the 
input value of the electromagnetic coupling $\alpha$. In fact,  
turning things around, one can use $a_e^\gamma$ to make the most precise 
determination of the fine structure constant \cite{KI:96,CKM:96}: 
\bel{eq:alpha} 
\alpha^{-1} = 137.03599959 \pm 0.00000040 \, . 
\ee 
The resulting accuracy is one order of magnitude better than 
the usually quoted value \cite{PDG:98} 
$\alpha^{-1} = 137.0359895 \pm 0.0000061$. 
 
\begin{figure}[tbh] 
\centering 
\vspace{0.3cm} 
\centerline{\epsfxsize =\linewidth \epsfbox{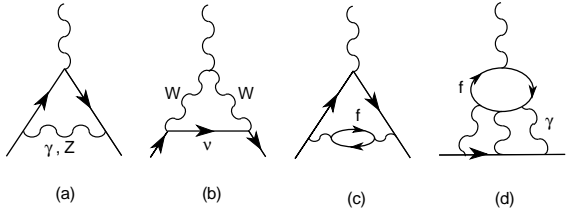}} 
\vspace{-0.5cm} 
\caption{Some Feynman diagrams contributing to $a_l^\gamma$. 
\label{fig:AnMagMom}} 
\end{figure} 

The anomalous magnetic moment of the muon is sensitive to virtual  
contributions from heavier states; compared to $a_e^\gamma$, they scale 
as $m_\mu^2/m_e^2$. 
The main theoretical uncertainty on $a_\mu^\gamma$ has a QCD origin. 
Since quarks have electric charge, virtual quark--antiquark pairs 
can be created by the photon leading to the so--called 
{\it hadronic vacuum polarization} corrections to the photon propagator 
(Figure 1.c). 
Owing to the non-perturbative character of QCD at low energies, 
the light--quark contribution cannot be reliably calculated at 
present; fortunately, this effect can be 
extracted from the measurement of the cross-section 
$\sigma(e^+e^-\to \mbox{\rm hadrons})$  at low energies, 
and from the invariant--mass distribution of the final hadrons in 
$\tau$ decays \cite{ADH:99}. 
The large uncertainties of the present data are the dominant 
limitation to the achievable theoretical precision on $a_\mu^\gamma$. 
It is expected that this will be improved at the DA$\Phi$NE 
$\Phi$ factory, where an accurate measurement of the hadronic production 
cross-section in the most relevant kinematical region is expected 
\cite{daphne}. 
Additional QCD uncertainties stem from the (smaller) 
{\it light--by--light scattering} 
contributions, where four photons couple to a light--quark loop 
(Figure 1.d); 
these corrections are under active investigation at present 
\cite{dRA:94,HK:97,BPP:96}. 
 
The improvement of the theoretical $a_\mu^\gamma$ prediction is of great interest 
in view of the new E821 experiment \cite{BNL:E821}, presently running 
at Brookhaven, 
which aims to reach a sensitivity of at least $4\times 10^{-10}$,  
and thereby observe the contributions from virtual $W^\pm$ and $Z$ bosons 
\cite{CKM:96,PPR:95,KKSS:92} 
($\delta a_\mu^\gamma|_{\mbox{\rms weak}} \sim 15 \times 10^{-10}$). 
The extent to which this measurement could provide a meaningful 
test of the electroweak theory depends critically on 
the accuracy one will be able to achieve pinning down the QCD corrections.

\section{LEPTONIC CHARGED--CURRENT COUPLINGS} 
\label{sec:cc} 
 
\begin{figure}[bht] 
\centerline{\epsfysize =2.7cm \epsfbox{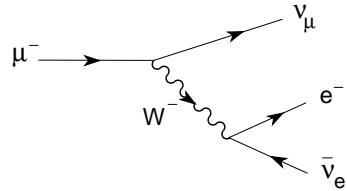}} 
\vspace{-0.5cm}
\caption{$\mu$--decay diagram.} 
\label{fig:mu_decay} 
\end{figure} 

The simplest flavour--changing process is the leptonic 
decay of the $\mu$, which proceeds through the $W$--exchange 
diagram shown in Figure~\ref{fig:mu_decay}. 
The momentum transfer carried by the intermediate $W$ is very small 
compared to $M_W$. Therefore, the vector--boson propagator reduces 
to a contact interaction. 
The decay can then be described through an effective local 
4--fermion Hamiltonian, 
\bel{eq:mu_v_a} 
\cH_{\mbox{\rms eff}}\, = \, {G_F \over\sqrt{2}} 
\left[\bar e\gamma^\alpha (1-\gamma_5) \nu_e\right]\, 
\left[ \bar\nu_\mu\gamma_\alpha (1-\gamma_5)\mu\right]\, ,  
\ee 
where 
\be 
{G_F\over\sqrt{2}} = {g^2\over 8 M_W^2}  
\ee 
is called the Fermi coupling constant.  
$G_F$ is fixed by the total decay width, 
\bel{eq:mu_lifetime} 
{1\over\tau_\mu} =    %
{G_F^2 m_\mu^5\over 192 \pi^3}\, 
\left( 1 + \delta_{\mbox{\rms RC}}\right) \,  
f\left(m_e^2/m_\mu^2\right) \, , 
\ee 
where 
$\, f(x) = 1-8x+8x^3-x^4-12x^2\ln{x}$, 
and 
$\delta_{\mbox{\rms RC}} = - 0.0042$  
takes into account the leading higher--order corrections 
\cite{KS:59,ST:99}. 
The measured $\mu$ lifetime \cite{PDG:98}, 
$\tau_\mu=(2.19703\pm 0.00004)\times 10^{-6}$~s, 
implies the value 
\beqn\label{eq:gf} 
G_F & = & (1.16637\pm 0.00001)\times 10^{-5} \:\mbox{\rm GeV}^{-2} 
\no\\ &\approx & (293 \:\mbox{\rm GeV})^{-2} \, . 
\eeqn 
%

\begin{figure}[tbh]
\centerline{\epsfysize =2.7cm \epsfbox{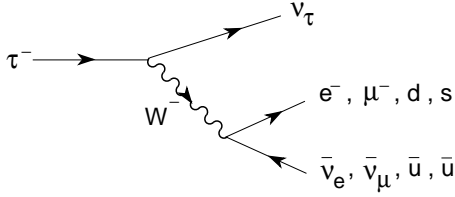}} 
\vspace{-0.5cm}
\caption{$\tau$--decay diagram.} 
\end{figure} 

The leptonic $\tau$ decay widths  
$\tau^-\to l^-\bar\nu_l\nu_\tau$ \ ($l=e,\mu$) 
are also given by Eq.~\eqn{eq:mu_lifetime}, 
making the appropriate changes for the masses of the initial and final 
leptons. 
Using the value of $G_F$   
measured in $\mu$ decay, one gets a relation between the $\tau$ lifetime 
and leptonic branching ratios \cite{tau96}: 
\beqn\label{eq:relation} 
B_{\tau\to e} &=&  {B_{\tau\to \mu} \over 0.972564\pm 0.000010}  
\no\\ &=&  
{ \tau_{\tau} \over (1.6321 \pm 0.0014) \times 10^{-12}\, \mbox{\rm s} } 
\, . 
\eeqn 
The errors reflect the present uncertainty of $0.3$ MeV 
in the value of $m_\tau$.

\begin{figure}[t]  
\centering 
\centerline{\epsfxsize =\linewidth \epsfbox{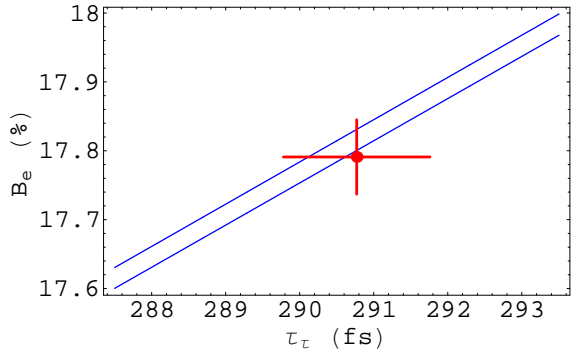}} 
\vspace{-0.5cm} 
\caption{Relation between $B_{\tau\to e}$ and $\tau_\tau$. The
band corresponds to the SM prediction in Eq.~(\protect\ref{eq:relation}). 
\label{fig:BeLife}} 
\end{figure} 
 
The measured ratio $B_{\tau\to\mu}/B_{\tau\to e} = 0.974 \pm 0.004$ 
is in perfect agreement with the predicted value. 
As shown in 
Figure~\ref{fig:BeLife}, the relation between $B_{\tau\to e}$ and 
$\tau_\tau$ is also well satisfied by the present data.  
The experimental precision (0.3\%) is already approaching the 
level where a possible non-zero $\nu_\tau$ mass could become relevant; the 
present bound \cite{tau96} 
$m_{\nu_\tau}< 18.2$ MeV (95\% CL) only guarantees that such  
effect is below 0.08\%. 

These measurements test the universality of the $W$ couplings 
to the leptonic charged currents. 
Allowing the coupling $g$ 
to depend on the considered lepton flavour  
(i.e.  $g_e$, $g_\mu$, $g_\tau$),  
the $B_{\tau\to\mu}/B_{\tau\to e}$ ratio  
constrains $|g_\mu/g_e|$, while $B_{\tau\to e}/\tau_\tau$ 
provides information on $|g_\tau/g_\mu|$. 
The present results \cite{tau96} 
are shown in Tables \ref{tab:univme}, 
\ref{tab:univtm} and \ref{tab:univte}, together with the values obtained  
from the ratios 
$R_{\pi\to e/\mu}\equiv\Gamma(\pi^-\to e^-\bar\nu_e)/ 
\Gamma(\pi^-\to \mu^-\bar\nu_\mu)$ 
and  
$R_{\tau/P} \equiv\Gamma(\tau^-\to\nu_\tau P^-) / 
 \Gamma(P^-\to \mu^-\bar\nu_\mu)\, $  [$P=\pi,K$], 
from the comparison of the $\sigma\cdot B$ partial production 
cross-sections for the various $W^-\to l^-\bar\nu_l$ decay 
modes at the $p\bar p$ colliders, 
and from the most recent LEP2 measurements of the leptonic 
$W^\pm$ branching ratios.  
 
\begin{table}[tbh] 
\centering 
\caption{Present constraints on $|g_\mu/g_e|$.} 
\label{tab:univme} 
\vspace{0.2cm} 
\begin{tabular}{|c|c|} 
\hline 
& $|g_\mu/g_e|$ \\ \hline 
$B_{\tau\to\mu}/B_{\tau\to e}$ & $1.0009\pm 0.0022$ 
\\ 
$B_{\pi\to e}/B_{\pi\to\mu}$ & $1.0017\pm 0.0015$ 
\\ 
$\sigma\cdot B_{W\to\mu/e}$ \ \ ($p\bar p$) & $0.98\pm 0.03$ 
\\ 
$B_{W\to\mu/e}$ (LEP2) & $1.002\pm 0.016$ 
\\ \hline 
\end{tabular}\vspace{0.3cm} 
%
\caption{Present constraints on $|g_\tau/g_\mu|$.} 
\label{tab:univtm} 
\vspace{0.2cm} 
\begin{tabular}{|c|c|} 
\hline 
& $|g_\tau/g_\mu|$  \\ \hline 
$B_{\tau\to e}\tau_\mu/\tau_\tau$ & $0.9993\pm 0.0023$ 
\\ 
$\Gamma_{\tau\to\pi}/\Gamma_{\pi\to\mu}$ &  $1.005\pm 0.005$ 
\\ 
$\Gamma_{\tau\to K}/\Gamma_{K\to\mu}$ & $0.981\pm 0.018$ 
\\ 
$B_{W\to\tau/\mu}$ (LEP2) & $1.008\pm 0.019$ 
\\ \hline 
\end{tabular}\vspace{0.3cm} 
%
\caption{Present constraints on $|g_\tau/g_e|$.} 
\label{tab:univte} 
\vspace{0.2cm} 
\begin{tabular}{|c|c|} 
\hline 
& $|g_\tau/g_e|$  \\ \hline 
$B_{\tau\to \mu}\tau_\mu/\tau_\tau$ & $1.0002\pm 0.0023$ 
\\ 
$\sigma\cdot B_{W\to\tau/e}$ \ \ ($p\bar p$) & $0.987\pm 0.025$ 
\\ 
$B_{W\to\tau/e}$ (LEP2) & $1.010\pm 0.019$ 
\\ \hline 
\end{tabular} 
\end{table} 
%

The present data verify the universality of the leptonic 
charged--current couplings to the 0.15\% ($\mu/e$) and 0.23\% 
($\tau/\mu$, $\tau/e$) level. The precision of the most recent 
$\tau$--decay measurements is becoming competitive with the  
more accurate $\pi$--decay determination.  
It is important to realize the complementarity of the 
different universality tests.  
The pure leptonic decay modes probe 
the charged--current couplings of a transverse $W$. In contrast, 
the decays $\pi/K\to l\bar\nu$ and $\tau\to\nu_\tau\pi/K$ are only 
sensitive to the spin--0 piece of the charged current; thus, 
they could unveil the presence of possible scalar--exchange 
contributions with Yukawa--like couplings proportional to some 
power of the charged--lepton mass.

\subsection{Lorentz Structure} 
\label{sec:lorentz}

Let us consider the leptonic 
decay $l^-\to\nu_l l'^-\bar\nu_{l'}$.  
The most general, local, derivative--free, lepton--number conserving,  
four--lepton interaction Hamiltonian,  
consistent with locality and Lorentz invariance 
\cite{MI:50,KS:57,FGJ:86,PS:95} 
\be 
{\cal H} = 4 \frac{G_{l'l}}{\sqrt{2}} 
\sum_{n,\epsilon,\omega}          
g^n_{\epsilon\omega}   
\left[ \overline{l'_\epsilon}  
\Gamma^n {(\nu_{l'})}_\sigma \right]\,  
\left[ \overline{({\nu_l})_\lambda} \Gamma_n  
	l_\omega \right]\ , 
\label{eq:hamiltonian} 
\ee 
contains ten complex coupling constants or, since a common phase is 
arbitrary, nineteen independent real parameters. 
The subindices 
$\epsilon , \omega , \sigma, \lambda$ label the chiralities (left--handed, 
right--handed)  of the  corresponding  fermions, and $n$ the 
type of interaction:  
scalar ($I$), vector ($\gamma^\mu$), tensor  
($\sigma^{\mu\nu}/\sqrt{2}$). 
For given $n, \epsilon , 
\omega $, the neutrino chiralities $\sigma $ and $\lambda$ 
are uniquely determined. 
Taking out a common factor $G_{l'l}$, which is determined by the total 
decay rate, the coupling constants $g^n_{\epsilon\omega}$ 
are normalized to \cite{FGJ:86} 
\bel{eq:normalization} 
1 = \sum_{n,\epsilon,\omega}\, |g^n_{\epsilon\omega}/N^n|^2 \, , 
\ee 
where  
$N^n 
=2$, 1, 
$1/\protect\sqrt{3} $ for $n =$ S, V, T. 
In the SM, $g^V_{LL}  = 1$  and all other 
$g^n_{\epsilon\omega} = 0 $.

\begin{figure}[tbh] 
\centerline{\epsfxsize =7cm \epsfbox{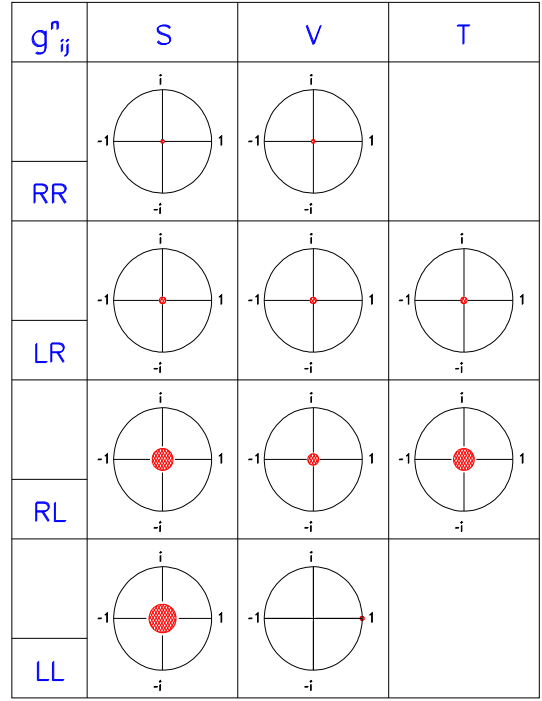}}
\vspace{-0.5cm}
\caption{90\% CL experimental limits  (shaded regions) \protect\cite{PDG:98} 
for the normalized $\mu$--decay couplings 
$g'^n_{\epsilon\omega }\equiv g^n_{\epsilon\omega }/ N^n$.}
\label{fig:mu_couplings} 
\end{figure}
\begin{figure}[bht]
\centerline{\epsfxsize =7cm \epsfbox{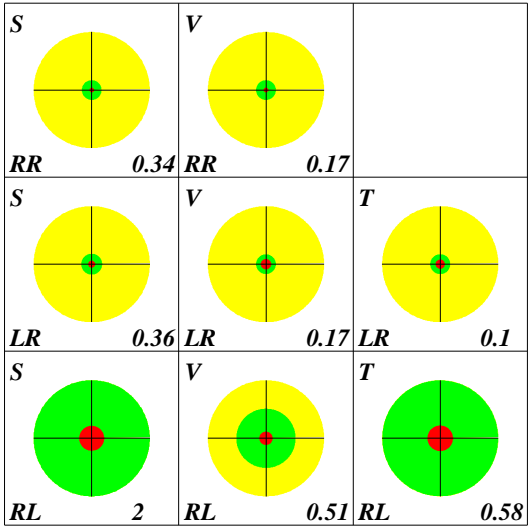}}
\vspace{-0.5cm}
\caption{90\% CL experimental limits \protect\cite{gtau99} 
for the normalized $\tau$--decay couplings 
$g'^n_{\epsilon\omega }\equiv g^n_{\epsilon\omega }/ N^n$, 
assuming $e/\mu$ universality. 
For comparison, the $\mu$--decay limits are also shown (darker circles). 
\hfill\hfil } 
\label{fig:tau_couplings} 
\end{figure} 

The couplings $g^n_{\epsilon\omega}$ can be investigated through the 
measurement of the final charged--lepton distribution  
and with the inverse decay 
$\nu_{l'} l\to l' \nu_l$.  
For $\mu$ decay, where precise measurements of the polarizations of 
both $\mu$ and $e$ have been performed,  
there exist \cite{PDG:98} 
stringent bounds on the couplings involving right--handed helicities. 
These limits show nicely that  
the $\mu$--decay transition amplitude is indeed of 
the predicted V$-$A type: 
$|g^V_{LL}|  > 0.96$ (90\% CL).

Figure \ref{fig:tau_couplings} shows the most recent limits 
on the $\tau$ couplings \cite{gtau99}. 
The circles of unit area indicate the range allowed by the
normalization constraint (\ref{eq:normalization}).
The present experimental bounds are shown as shaded circles.
For comparison, the (stronger) $\mu$-decay limits are also
given (darker circles).
The measurement of the $\tau$ polarization allows to bound those couplings 
involving an initial right--handed lepton; however, information on the 
final charged--lepton polarization is still lacking. 
The measurement of the inverse decay 
$\nu_\tau l\to\tau\nu_l$, needed to separate the $g^S_{LL}$ and 
$g^V_{LL}$ couplings, looks far out of reach.

\section{NEUTRAL--CURRENT COUPLINGS} 
\label{sec:nc} 
 
In the SM, all fermions with equal electric charge have identical vector,  
$v_f = T_3^f (1-4|Q_f|\sin^2{\theta_W})$ and axial--vector, 
$a_f=T_3^f$, couplings to the $Z$ boson. 
These neutral current couplings have been precisely 
tested at LEP and SLC. 
 
The gauge sector of the SM is fully described in terms of only 
four parameters: $g$, $g'$,  
and the two constants characterizing the scalar potential. 
We can trade these parameters by \cite{jaca:94,lepewwg:97,swartz} 
$\alpha$, $G_F$, 
\bel{eq:SM_inputs} 
M_Z  =  (91.1871\pm 0.0021)\,\mbox{\rm GeV} \, ,  
\ee 
and $m_H$; this has the advantage of using the 3  
most precise experimental determinations to fix the interaction. 
The relations 
\bel{eq:A_def} 
M_W^2 s_W^2  =  {\pi\alpha\over\sqrt{2} G_F}\, , 
\qquad\quad 
s_W^2  =  1 - {M_W^2\over M_Z^2}\, , 
\ee 
determine then $s_W^2 \equiv \sin^2{\theta_W} = 0.2122$  
and $M_W = 80.94$ GeV; 
in reasonable agreement 
with the measured $W$ mass \cite{lepewwg:97,swartz},  
$M_W = 80.394\pm 0.042$ GeV. 
 
At tree level, the partial decay widths of the $Z$ boson  
are given by 
%
\bel{eq:Z_width} 
\Gamma\left[ Z\to \bar f f\right]  =   
{G_F M_Z^3\over 6\pi\sqrt{2}} \, \left(|v_f|^2 + |a_f|^2\right)\,  
N_f   
\, , 
\ee 
where $N_l=1$ and $N_q=N_C$. 
Summing over all possible final fermion pairs, one predicts 
the total width 
$\Gamma_Z=2.474$ GeV, to be compared 
with the experimental value \cite{lepewwg:97,swartz} 
$\Gamma_Z=(2.4944\pm 0.0024)$ GeV. 
The leptonic decay widths of the $Z$ are predicted to be  
$\Gamma_l\equiv\Gamma(Z\to l^+l^-) =  
84.84$ MeV, 
in agreement with the measured value 
$\Gamma_l = (83.96\pm 0.09)$ MeV.

Other interesting quantities are  
the ratios 
$R_l\equiv\Gamma(Z\to\mbox{\rm hadrons})/\Gamma_l$ 
and 
$R_Q\equiv\Gamma(Z\to\bar Q Q)/ \Gamma(Z\to\mbox{\rm hadrons})$. 
The comparison between the tree--level theoretical predictions  
and the experimental values, shown  in 
Table~\ref{tab:results}, is quite good. 
 
 
Additional information can be obtained from the study of the 
fermion--pair production process 
$e^+e^-\to\gamma,Z\to\bar f f $. 
LEP has provided accurate measurements of 
the total cross-section, the forward--backward asymmetry, 
the polarization asymmetry and the forward--backward polarization 
asymmetry, at the $Z$ peak ($s=M_Z^2$): 
%
\beqn 
\sigma^{0,f}  
 = {12 \pi  \over M_Z^2 } \, {\Gamma_e \Gamma_f\over\Gamma_Z^2}\, , 
&& \quad 
\cA_{\mbox{\rms FB}}^{0,f}  
= {3 \over 4} \cP_e \cP_f \, , 
\no\\ \label{eq:A_pol_Z} 
\cA_{\mbox{\rms Pol}}^{0,f}  
= \cP_f \, ,\qquad\quad 
&&  \quad 
\cA_{\mbox{\rms FB,Pol}}^{0,f}  
=  {3 \over 4} \cP_e  \, ,\quad\quad 
\eeqn 
where  
$\Gamma_f$ is the $Z$ partial decay width to the $\bar f f$ final state, and 
\bel{eq:P_f} 
\cP_f \, \equiv \, { - 2 v_f a_f \over v_f^2 + a_f^2}  
\ee 
is the average longitudinal polarization of the fermion $f$. 

The measurement of the final polarization asymmetries can (only) be done for  
$f=\tau$, because the spin polarization of the $\tau$'s 
is reflected in the distorted distribution of their decay products. 
Therefore, $\cP_\tau$ and $\cP_e$ can be determined from a 
measurement of the spectrum of the final charged particles in the 
decay of one $\tau$, or by studying the correlated distributions 
between the final products of both $\tau's$ \cite{ABGPR:92}. 
 
With polarized $e^+e^-$ beams, one can also study the left--right 
asymmetry between the cross-sections for initial left-- and right--handed 
electrons. 
At the $Z$ peak, this asymmetry directly measures  
the average initial lepton polarization, $\cP_e$, 
without any need for final particle identification. 
SLD has also measured the left--right forward--backward asymmetries, 
which are only sensitive to the final state couplings: 

\bel{eq:A_LR} 
\cA_{\mbox{\rms LR}}^0   
 =  - \cP_e \,  , 
\qquad 
\cA_{\mbox{\rms FB,LR}}^{0,f} 
=  -{3\over 4} \cP_f \, . 
\ee 
%

\begin{table*}[thb] 
\begin{center} 
\caption{Comparison between SM predictions and 
experimental \protect\cite{lepewwg:97,swartz} measurements.  
The third column includes 
the main QED and QCD corrections. 
The experimental value for $s_W^2$ refers to the 
effective electroweak mixing angle in the charged--lepton sector. 
\hfill  
\label{tab:results}} 
\vspace{0.2cm} 
\begin{tabular}{|c|c|c|c|c|c|} 
\hline 
Parameter & \multicolumn{2}{c|}{Tree--level prediction} &  
SM fit & Experimental   & Pull 
\\ \cline{2-3} & Naive & Improved & (1--loop) & value & 
$\left(x_{\mbox{\rms Exp}} - x_{\mbox{\rms fit}}\right)/
\sigma_{\mbox{\rms Exp}}$
\\ \hline  
$M_W$ \, (GeV) & 80.94 & 79.96 & $80.385$ & $80.394\pm 0.042$ & $0.21$
\\ 
$s_W^2$ & 0.2122 & 0.2311 & $0.23150$ & $0.23153\pm 0.00017$ & $0.18$
\\ 
$\Gamma_Z$ \, (GeV) & 2.474 & 2.490 & $2.4957$ & $2.4944\pm 0.0024$
& $-0.56$ 
\\ 
$R_l$ & 20.29 & 20.88 & 20.740 & $20.768\pm 0.024$ & $1.16$
\\ 
$\sigma^0_{\mbox{\rms had}}$ \,\, (nb) & 42.13 & 41.38 & 41.479 & 
$41.544\pm 0.037$ & $1.75$
\\ 
$\cA_{\mbox{\rms FB}}^{0,l}$ & 0.0657 & 0.0169 & 0.01625 & 
$0.01701\pm 0.00095$  & $0.80$
\\ 
$\cP_l$ & $-0.296$ & $-0.150$ & $-0.1472$ & $-0.1497\pm 0.0016$ & $-1.56$
\\ 
$\cA_{\mbox{\rms FB}}^{0,b}$ & 0.210 & 0.105 & 0.1032 & $0.0988\pm 0.0020$ 
& $-2.20$
\\ 
$\cA_{\mbox{\rms FB}}^{0,c}$ & 0.162 & 0.075 & 0.0738 & $0.0692\pm 0.0037$ 
& $-1.23$
\\ 
$\cP_b$ & $-0.947$ & $-0.936$ & $-0.935$ & $-0.905\pm 0.026$ & $1.15$
\\ 
$\cP_c$ & $-0.731$ & $-0.669$ & $-0.668$ & $-0.634\pm 0.027$ & $1.26$
\\ 
$R_b$ & 0.219 & 0.220 & 0.21583 & $0.21642\pm 0.00073$ & $0.81$ 
\\ 
$R_c$ & 0.172 & 0.170 & 0.1722 & $0.1674\pm 0.0038$ & $-1.27$
\\ \hline 
\end{tabular} 
\end{center} 
\end{table*} 

Using $s_W^2 = 0.2122$, 
one gets the (tree--level) predictions shown in the second column of  
Table~\ref{tab:results}. 
The comparison with the experimental measurements 
looks reasonable for the total hadronic cross-section  
$\sigma^0_{\mbox{\rms had}} \equiv \sum_q \, \sigma^{0,q}$; 
however, all leptonic asymmetries disagree with 
the measured values by several standard deviations. 
As shown in the table, the same happens with the  
heavy--flavour forward--backward asymmetries 
$\cA_{\mbox{\rms FB}}^{0,b/c}$, 
which compare very badly with the experimental measurements; 
the agreement is however better for $\cP_{b/c}$. 
 
Clearly, the problem with the asymmetries is their high sensitivity 
to the input value of $\sin^2{\theta_W}$; 
specially the ones involving the leptonic vector coupling 
$v_l = (1 - 4 \sin^2{\theta_W})/2$. Therefore, they are an 
extremely good window into higher--order electroweak corrections.

\subsection{Important QED and QCD Corrections} 
\label{subsec:QED_QCD_corr} 
 
The photon propagator gets vacuum polarization corrections, induced by 
virtual fermion--antifermion pairs. Their effect can be 
taken into account through a redefinition of the QED coupling, 
which depends on the energy scale of the process; 
the resulting effective coupling $\alpha(s)$ 
is called the QED {\it running coupling}. 
The fine structure constant is measured 
at very low energies; it corresponds to $\alpha(m_e^2)$. 
However, at the $Z$ peak, we should rather use $\alpha(M_Z^2)$. 
The long running from $m_e$ to $M_Z$ gives rise to a sizeable 
correction \cite{ADH:99,EJ:95}: 
$\alpha(M_Z^2)^{-1} =   
128.878\pm  0.090\, $. 
The quoted uncertainty arises from the light--quark contribution, 
which is estimated from $\sigma(e^+e^-\to\mbox{\rm hadrons})$ and  
$\tau$--decay data.  
 
Since $G_F$ is measured at low energies, while $M_W$ is a 
high--energy parameter, the relation between both quantities  
in Eq.~\eqn{eq:A_def} 
is clearly modified by vacuum--polarization contributions. 
One gets then the corrected predictions 
$M_W = 79.96$ GeV and $s^2_W = 0.2311$. 
 
The gluonic corrections to the $Z\to\bar q q$ decays  
can be directly incorporated 
by taking an effective number of colours 
$N_q = N_C\,\left\{ 1 + {\alpha_s\over\pi} + \ldots\right\}\, 
\approx\, 3.12$, 
where we have used $\alpha_s(M_Z^2)\approx 0.12\, $. 

The third column in Table~\ref{tab:results} shows the numerical impact 
of these  QED and QCD corrections. In all cases, the comparison with 
the data gets improved. However, it is in the asymmetries where the 
effect gets more spectacular. Owing to the high sensitivity to $s^2_W$, 
the small change in the value of the weak mixing angle generates 
a huge difference of about a factor of 2 in the predicted 
asymmetries. 
The agreement with the experimental values is now very good.

\subsection{Higher--Order Electroweak Corrections} 
\label{subsec:nc-loop} 
 
 Initial-- and final--state photon radiation is by far the 
most important numerical correction. One has in addition the contributions 
coming from photon exchange between the fermionic lines.  
All these QED corrections are to a large extent dependent on the detector and 
the experimental cuts, because of the infra-red problems associated with 
massless photons. 
 (one needs to define, for instance, the minimun photon energy 
which can be detected).  
These effects are usually estimated with 
Monte Carlo programs and subtracted from the data. 

More interesting are the so--called {\it oblique} corrections, 
gauge--boson self-energies induced by vacuum polarization diagrams, 
which are {\it universal} (process independent).  
In the case of the $W^\pm$ and the $Z$, these corrections are sensitive 
to heavy particles (such as the top) running along the loop \cite{VE:77}. 
In QED, the  
vacuum polarization contribution of a heavy fermion pair 
is suppressed by inverse powers of the fermion mass. 
At low energies ($s<<m_f^2$),  
the information on the heavy fermions is then lost. 
This {\it decoupling} of the heavy fields happens in theories 
like QED and QCD, 
with only vector couplings and an exact gauge symmetry \cite{AC:75}. 
The SM involves, however, a broken chiral gauge symmetry.  
%
The $W^\pm$ and $Z$ self-energies induced by a heavy top 
generate contributions  
which increase quadratically with the top mass \cite{VE:77}. 
The leading $m_t^2$ contribution to the $W^\pm$ propagator  
amounts to a $-3\% $  correction to the relation \eqn{eq:A_def} 
between $G_F$ and $M_W$. 
 
Owing to an accidental $SU(2)_C$ symmetry of the scalar sector, 
the virtual production of Higgs particles does not generate any 
$m_H^2$ dependence at one loop \cite{VE:77}. 
The dependence on the Higgs mass is only logarithmic. 
The numerical size of the correction induced on \eqn{eq:A_def} 
is $-0.3\% $ ($+1\% $) for $m_H=60$ (1000) GeV. 

\begin{figure}[tbh]
\centerline{\epsfxsize =\linewidth \epsfbox{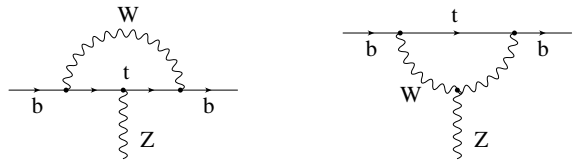}}
\vspace{-0.5cm}
\caption{$m_t$--dependent corrections to the $Z\bar b b$ vertex.
\hfill\hfil }
\label{fig:zbb}
\end{figure}

The vertex corrections 
are {\it non-universal} and usually smaller than the oblique contributions. 
There is one interesting exception, the $Z \bar bb$ vertex, which is sensitive 
to the top quark mass \cite{BPS:88}. 
The $Z\bar f f$ vertex gets 1--loop corrections where a virtual 
$W^\pm$ is exchanged between the two fermionic legs.  
Since, the $W^\pm$ coupling changes the fermion flavour,  
the decays 
$Z\to \bar d_i\bar d_i$   
get contributions with a top quark 
in the internal fermionic lines. 
These amplitudes are suppressed by a small quark--mixing factor  
$|V_{td_i}|^2$, 
except for the $Z\to\bar b b$ vertex because $|V_{tb}|\approx 1$. 
The explicit calculation \cite{BPS:88,ABR:86}    
shows the presence of hard $m_t^2$ corrections to 
the $Z\to\bar b b$ vertex, 
which amount to a  $-1.5\% $ effect in $\Gamma(Z\to\bar b b)$.

The {\it non-decoupling} present in the 
$Z\bar b b$ vertex is quite different from the one happening in 
the boson self-energies.  
The vertex correction does not have any dependence with the 
Higgs mass. Moreover, 
while any kind of new heavy particle, 
coupling to the gauge bosons, would contribute to the $W^\pm$ and $Z$ 
self-energies, possible new--physics contributions to the 
$Z\bar b b$ vertex are much more restricted and, in any case, 
different. 
Therefore, an independent experimental test of the two effects 
is very valuable in order to disentangle possible 
new--physics contributions from the SM corrections.

The remaining quantum corrections (box diagrams, Higgs exchange) 
are rather small at the $Z$ peak.

\subsection{Lepton Universality} 
 
\begin{table}[tbh] 
\centering 
\caption{Measured values \protect\cite{lepewwg:97,swartz} 
of $\Gamma_l$  
and the leptonic forward--backward asymmetries. 
The last row shows the combined result  
(for a massless lepton) assuming lepton universality. \hfill 
\label{tab:LEP_asym}} 
\vspace{0.2cm} 
\begin{tabular}{|c|c|c|} 
\hline 
& $\Gamma_l$ \, (MeV) &  $\cA_{\mbox{\rms FB}}^{0,l}$ \, (\%) 
\\ \hline 
$e$ & $83.90\pm 0.12$ & $1.45\pm 0.24$ 
\\ 
$\mu$ & $83.96\pm 0.18$ & $1.67\pm 0.13$ 
\\ 
$\tau$ & $84.05\pm 0.22$ & $1.88\pm 0.17$ 
\\ \hline 
$l$ & $83.96\pm 0.09$ & $1.701\pm 0.095$ 
\\ \hline 
\end{tabular}\vspace{0.2cm} 
\centering 
\caption{Measured values \protect\cite{lepewwg:97,swartz} 
of the leptonic polarization asymmetries.} 
\label{tab:pol_asym} 
\vspace{0.2cm} 
\begin{tabular}{|c|c|} 
\hline 
$-\cA_{\mbox{\rms Pol}}^{0,\tau} = -\cP_\tau$ & $0.1425\pm 0.0044$ 
\\ 
$-{4\over 3}\cA^{0,\tau}_{\mbox{\rms FB,Pol}} = -\cP_e$ & 
$0.1483\pm 0.0051$ 
\\ 
$\cA_{\mbox{\rms LR}}^0 = -\cP_e$ & $0.1511\pm 0.0022$ 
\\  
$\!\{{4\over 3}\cA_{\mbox{\rms FB}}^{0,l}\}^{1/2} = -P_l\! $ 
& $0.1506\pm 0.0042$ 
\\ \hline 
${4\over 3}\cA^{0,e}_{\mbox{\rms FB,LR}} \Rightarrow -\cP_e$ & 
$0.1558\pm 0.0064$ 
\\ 
${4\over 3}\cA^{0,\mu}_{\mbox{\rms FB,LR}} = -\cP_\mu$ & 
$0.137\pm 0.016$ 
\\ 
${4\over 3}\cA^{0,\tau}_{\mbox{\rms FB,LR}} = -\cP_\tau$ & 
$0.142\pm 0.016$ 
\\ \hline
\end{tabular} 
\end{table} 

\begin{figure}[tbh]  
\centerline{\epsfxsize =\linewidth \epsfbox{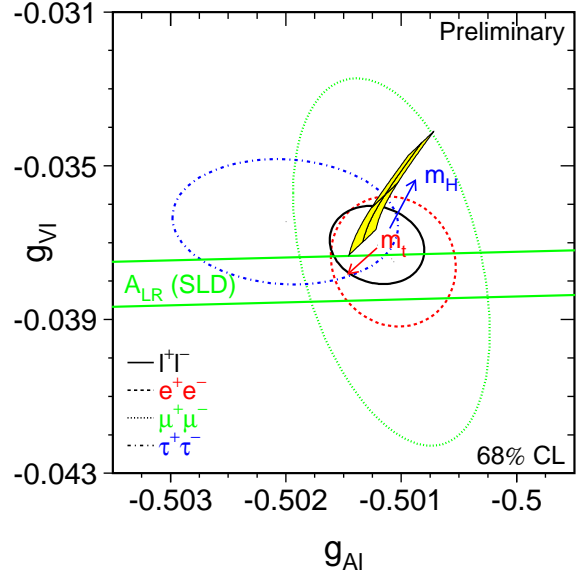}} 
\vspace{-0.6cm} 
\caption{68\% probability contours in the $a_l$-$v_l$ plane 
from LEP measurements \protect\cite{lepewwg:97}.  
The solid contour assumes lepton universality.  
Also shown is the $1\sigma$ band resulting from the 
$\protect\cA_{\mbox{\protect\rms LR}}^0$ measurement at SLD.  
The shaded region corresponds to the SM prediction 
for \protect{$m_t= 174.3\pm 5.1$} GeV and 
\protect{$m_H = 300^{+700}_{-210}$} GeV. 
The arrows point in the direction of increasing 
$m_t$ and $m_H$ values.  
\hfill\hfil } 
\label{fig:gagv} 
\end{figure} 

Tables~\ref{tab:LEP_asym} and \ref{tab:pol_asym} 
show the present experimental results 
for the leptonic $Z$ decay widths and asymmetries. 
The data are in excellent agreement with the SM predictions 
and confirm the universality of the leptonic neutral couplings. 
The average of the two $\tau$ polarization measurements, 
$\cA_{\mbox{\rms Pol}}^{0,\tau}$ and 
${4\over 3}\cA^{0,\tau}_{\mbox{\rms FB,Pol}}$, 
results in $\cP_l = -0.1450\pm 0.0033$ which deviates
by $1.5\,\sigma$ from the $\cA^0_{LR}$ measurement. 
Assuming lepton universality,  
the combined result from all leptonic asymmetries gives 
\bel{eq:average_P_l} 
\cP_l = - 0.1497\pm 0.0016  \, .
\ee 

Figure~\ref{fig:gagv} shows the 
68\% probability contours in the $a_l$--$v_l$ plane, 
obtained from  a combined analysis \cite{lepewwg:97}  
of all leptonic observables.  
Lepton universality is now tested to the $0.15\% $ level for the
axial--vector neutral couplings, while only a few per cent precision
has been achieved for the vector couplings \cite{swartz}:
\beqn
{a_\mu \over a_e} = 1.0001\pm 0.0014 & , &
{v_\mu \over v_e} = 0.981\pm 0.082\, ,
\no\\
{a_\tau \over a_e} = 1.0019\pm 0.0015 & , &
{v_\tau \over v_e} = 0.964\pm 0.032 \, .
\no\eeqn

The neutrino couplings can be determined from the invisible  
$Z$--decay width,  
$\Gamma_{\mbox{\rms inv}}/\Gamma_l = 5.941\pm 0.016$, 
by assuming three identical neutrino generations 
with left--handed couplings  
and fixing the sign from neutrino scattering  
data \cite{CHARMII:94}. 
The resulting experimental value \cite{lepewwg:97}, 
$v_\nu=a_\nu = 0.50123\pm 0.00095$, 
is in perfect agreement with the SM. 
Alternatively, one can use the SM prediction,  
$\Gamma_{\mbox{\rms inv}}/\Gamma_l = (1.9912\pm 0.0012) \, N_\nu$, 
to get a determination of the number of (light) neutrino flavours 
\cite{lepewwg:97,swartz}: 
\be 
N_\nu = 2.9835\pm 0.0083 \, . 
\ee 
The universality of the neutrino couplings has been tested 
with $\nu_\mu e$ scattering data, which fixes \cite{CHARMII:94b} 
the $\nu_\mu$ coupling to the $Z$: \  
$v_{\nu_\mu} =  a_{\nu_\mu} = 0.502\pm 0.017$. 
 
Assuming lepton universality, 
the measured leptonic asymmetries can be used to obtain the 
effective electroweak mixing angle in the charged--lepton sector 
($\chi^2/\mbox{d.o.f.} = 3.4/4$):
$$ 
\sin^2{\theta^{\mbox{\rms lept}}_{\mbox{\rms eff}}} \equiv 
{1\over 4}   \left( 1 - {v_l\over a_l}\right)  
\, = \, 0.23119\pm 0.00021
\, . 
$$ 
%
Including also the information provided by the hadronic 
asymmetries, one gets \cite{lepewwg:97,swartz}  
$\sin^2{\theta^{\mbox{\rms lept}}_{\mbox{\rms eff}}} = 0.23153\pm 0.00017$,
with a $\chi^2/\mbox{d.o.f.} = 13.3/7$.

\subsection{SM Electroweak Fit}

\begin{figure}[tbh]
\centerline{\epsfxsize =\linewidth \epsfbox{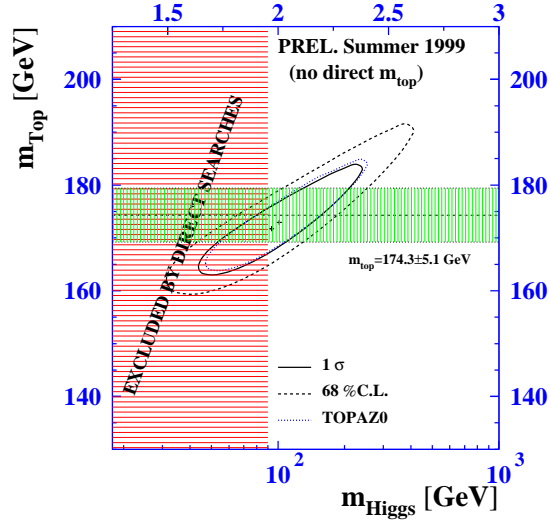}}
\vspace{-0.7cm}
\caption{Contours in $m_t$ and $m_H$ obtained from the SM electroweak fit.
Also shown are the 95\% exclusion limit on $m_H$ from direct searches and
the Fermilab measurement of $m_t$
\protect\cite{lepewwg:97}.
\hfill\hfil }
\label{fig:mhmt}
\end{figure}

The high accuracy of the present data provides compelling evidence 
for the pure weak quantum corrections, beyond the main QED and QCD corrections 
discussed in Section~\ref{subsec:QED_QCD_corr}. 
The measurements are sufficiently precise to require the presence of 
quantum corrections associated with the  
virtual exchange of top quarks, gauge bosons and Higgses.

Figure \ref{fig:mhmt} shows the constraints obtained on $m_t$ and $m_H$,
from a global fit to the electroweak data \cite{lepewwg:97}.
The fitted value of the top mass is in excellent agreement with
the direct Tevatron measurement $m_t = 174.3\pm 5.1$ GeV \cite{lepewwg:97}.
The data prefers a light Higgs, close to the present lower bound from
direct searches, $m_H > 95.2$ GeV (95\% CL). 
There is a large correlation between the fitted values of $m_t$ 
and $m_H$; the correlation would be much larger if the $R_b$ measurement 
was not used ($R_b$ is insensitive to $m_H$). 
The fit gives the upper bound \cite{lepewwg:97}: 
\bel{eq:M_H} 
m_H < 245 \;\mbox{\rm GeV} \qquad (95\% \, \mbox{\rm CL}) \, . 
\ee 

The global fit results
in an extracted value of the strong coupling, 
$\alpha_s(M_Z^2) = 0.119\pm 0.003$,
which agrees very well 
with the world average value \cite{PDG:98} $\alpha_s(M_Z^2) = 0.119\pm 
0.002$. 

As shown in Table \ref{tab:results}, the different electroweak measurements
are well reproduced by the SM electroweak fit. At present, the larger
deviation appears in  $\cA_{\mbox{\rms FB}}^{0,b}$, which seems to be
too low by $2.2 \,\sigma$.

The uncertainty on the QED coupling $\alpha(M_Z^2)^{-1}$ introduces 
a severe limitation on the accuracy of the SM predictions.
The uncertainty of the ``standard'' value, 
$\alpha(M_Z^2)^{-1} = 128.878\pm 0.090$ \cite{EJ:95},
causes an error of $0.00023$ on the
$\sin^2{\theta^{\mbox{\rms lept}}_{\mbox{\rms eff}}}$ prediction.
A recent analysis  \cite{ADH:99}, 
using hadronic $\tau$--decay data, results in a more
precise value,
$\alpha(M_Z^2)^{-1} = 128.933\pm 0.021$,
reducing the corresponding uncertainty on
$\sin^2{\theta^{\mbox{\rms lept}}_{\mbox{\rms eff}}}$
to $5\times 10^{-4}$;
this translates into a $30\% $ reduction in the error
of the fitted $\log{\left( m_H\right) }$ value.

To improve the present determination of $\alpha(M_Z^2)^{-1}$ 
one needs to perform a good measurement of 
$\sigma(e^+e^-\to \mbox{\rm hadrons})$, as a function of the centre--of--mass 
energy, in the whole kinematical range spanned by DA$\Phi$NE, a  
tau--charm factory and the B factories. 
This would result in a much stronger constraint on the Higgs mass.

\section{SUMMARY}
\label{sec:summary}

The SM provides a beautiful theoretical framework which is able to
accommodate all our present knowledge on electroweak interactions.
It is able to explain any single experimental fact  and, in some cases,
it has successfully passed very precise
tests at the 0.1\% to 1\% level. 
However, there are still  pieces of the SM Lagrangian which so far
have not been experimentally analyzed in any precise way.

The gauge self-couplings are presently being investigated at LEP2, through the
study of the $e^+e^-\to W^+W^-$ production cross-section.
The $V-A$ ($\nu_e$-exchange in the $t$ channel) contribution generates
an unphysical growing of the  cross-section with the centre-of-mass energy, 
which is compensated through a delicate gauge cancellation with the
$e^+e^-\to\gamma, Z\to W^+W^-$ amplitudes.
The recent LEP2 measurements of $\sigma(e^+e^-\to W^+W^-)$,
in good agreement with the SM,
have provided already convincing evidence \cite{lepewwg:97}
for the contribution coming from the $ZWW$ vertex.

The study of this process has also provided a more accurate measurement of
$M_W$, allowing to improve the precision of the neutral--current analyses.
The present LEP2 determination,
$M_W = 80.350\pm 0.056$ GeV, is already more precise than the value
$M_W = 80.448\pm 0.062$ GeV obtained in $p\bar p$ colliders.
Moreover it is in nice agreement with the result 
$M_W = 80.364\pm 0.029$ GeV
obtained from the indirect SM fit of electroweak data
\cite{lepewwg:97}.

The Higgs particle is the main missing block of the SM framework.
The data provide a clear confirmation of the assumed pattern of
spontaneous symmetry breaking, but do not prove the minimal Higgs
mechanism embedded in the SM.
At present, a relatively light Higgs is preferred by the
indirect precision tests.
LHC will try to find out whether such scalar field exists.

In spite of its enormous phenomenological success, the SM leaves too many
unanswered questions to be considered as a complete description of the
fundamental forces.
We do not understand yet why fermions are replicated in three
(and only three)
nearly identical copies? Why the pattern of masses and mixings
is what it is?  Are the masses the only difference among the three
families? What is the origin of the SM flavour structure?
Which dynamics is responsible for the observed CP violation?

Clearly, we need more experiments in order to learn what kind of
physics exists beyond the present SM frontiers.
We have, fortunately, a very promising and exciting future
ahead of us.

\vspace{0.5cm} 
 
This work has been supported in part by 
the ECC, TMR Network $EURODAPHNE$
 (ERBFMX-CT98-0169), and by
DGESIC (Spain) under grant No. PB97-1261.


\end{document}